\documentclass[10pt,twoside]{Lowx2011}
\usepackage{epsf,amsmath}
\usepackage{graphicx}

\begin{document}

\title{CENTRAL PRODUCTION OF GLUONIC AND QUARK-ANTIQUARK DIJETS
AND BACKGROUND TO CENTRAL PRODUCTION OF HIGGS BOSON
\thanks{This work was partially supported by the polish grant
MNiSW N N20224900235.}}

\author{\underline{A. SZCZUREK} \\ \\
Rzesz\'ow University, Rzesz\'ow, Poland and\\
Institute of Nuclear Physics, Krak\'ow, Poland\\
E-mail: antoni.szczurek@ifj.edu.pl}

\maketitle

\begin{abstract}
\noindent
We discuss exclusive central production of Higgs boson,
quark-antiquark and digluon dijets. Several differential distributions
are shown and disussed.
Irreducible leading-order $b \bar b $ background to Higgs production is 
calculated in several kinematical variables.
The signal-to-background ratio is shown and several improvements are 
suggested by imposing cuts on $b$ ($\bar b$) transverse momenta
and rapidities.
\end{abstract}



\markboth{\large \sl \hspace*{0.25cm}\underline{Antoni Szczurek} 
\hspace*{0.25cm} Low-$x$ Meeting 2011} {\large \sl \hspace*{0.25cm} TEMPLATE FOR THE
LOW$x$ 2011 MEETING PROCEEDINGS}

\section{Introduction}

Some time ago Khoze, Martin and Ryskin developed a QCD approach for
exclusive production of Higgs boson \cite{KMR_Higgs}.
The approach can be easily generalized to other exclusive
processes.
Recently we have applied this approach to Standard Model Higgs boson, 
quark-antiquark and digluon exclusive production
\cite{MPS10ccbar,MPS10higgs,MPS11higgs,MPS2011_gluons}.

Since the cross section for exclusive Higgs boson production is rather
small, only $b \bar b$ final state can be used to identify Higgs
boson. This means that a $b \bar b$ continuum background is of crucial
importance. Here we discuss this irreducible background.

In our calculations we include exact matrix elements
and perform full three- or four-body calculations for all considered 
processes. 
The kinematically complete calculations allow to include cuts on 
kinematical variables which is very usefull in order to identify
the Higgs boson signal.

\section{Formalism}

\begin{figure}
\includegraphics[width=6.0cm]{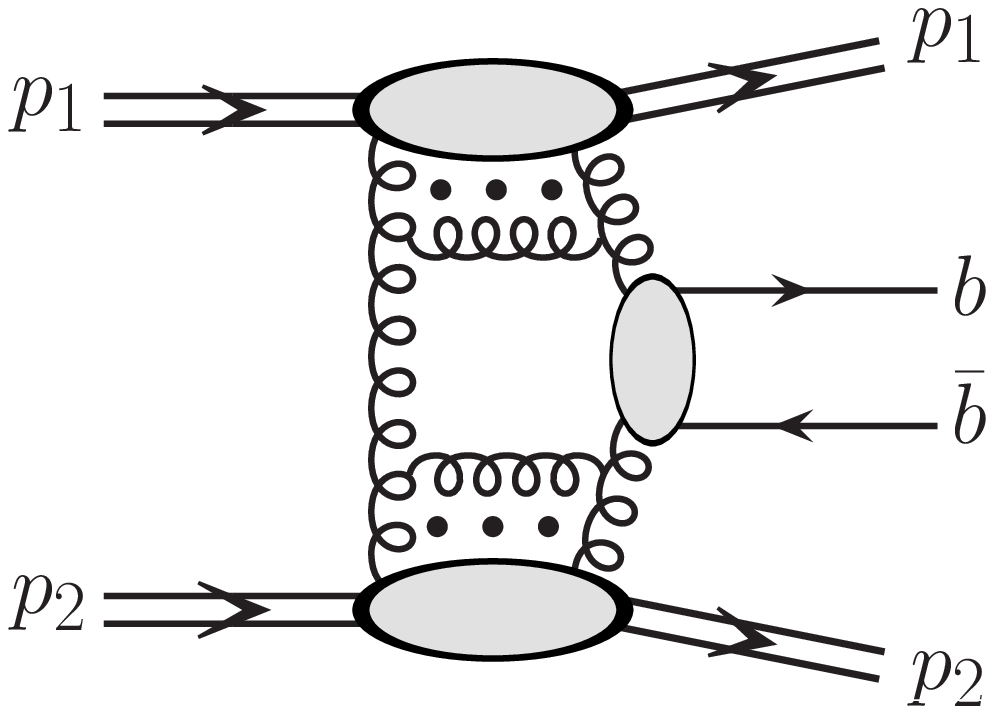}
\includegraphics[width=6.0cm]{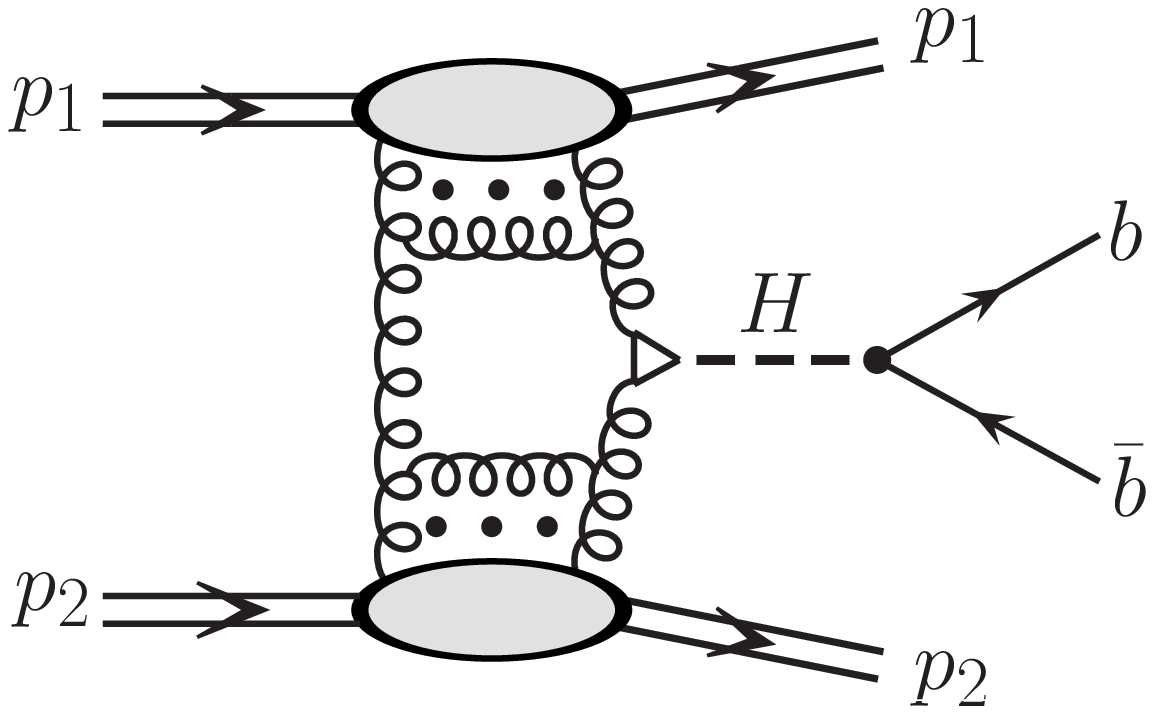}
\caption{The diagrams for the Higgs and background production.}
\label{fig:bbbar_diagrams}
\end{figure}

Let us concentrate on the simplest case of the production of $q\bar{q}$
pair in the color singlet state (see Fig.\ref{fig:bbbar_diagrams}). 
Color octet state would demand an
emission of an extra gluon which considerably complicates the
calculations. We do not consider the $q \bar q g$
contribution as it is higher order compared to the one considered here.

We write here only the amplitude of the exclusive diffractive $q\bar{q}$ pair
production $pp\to p(q\bar{q})p$ in the color singlet state as
\footnote{It is strightforward to write the amplitude for the other
  processes considered here.}
\begin{equation}
\begin{array}{lll}
{\cal M}_{\lambda_q\lambda_{\bar{q}}}^{p p \to p p q \bar q}(p'_1,p'_2,k_1,k_2) =
s\cdot\pi^2\frac12\frac{\delta_{c_1c_2}}{N_c^2-1}\, 
\Im \int d^2
q_{0,t} \; V_{\lambda_q\lambda_{\bar{q}}}^{c_1c_2}(q_1, q_2, k_1, k_2) \\
 \frac{f^{\mathrm{off}}_{g,1}(x_1,x_1',q_{0,t}^2,
q_{1,t}^2,t_1) \; f^{\mathrm{off}}_{g,2}(x_2,x_2',q_{0,t}^2,q_{2,t}^2,t_2)}
{q_{0,t}^2\,q_{1,t}^2\, q_{2,t}^2} \; ,
\end{array}
\label{amplitude}
\end{equation}
where $\lambda_q,\,\lambda_{\bar{q}}$ are helicities of heavy $q$
and $\bar{q}$, respectively. Above $f_1^{\mathrm{off}}$ and
$f_2^{\mathrm{off}}$ are the off-diagonal unintegrated gluon
distributions in nucleon 1 and 2, respectively. 
The longitudinal momentum fractions of active gluons
are calculated based on kinematical variables of outgoing quark
and antiquark.
The bare amplitude above is subjected to absorption corrections.
The absorption corrections are taken here in a multiplicative form.

The color singlet $q\bar{q}$ pair production amplitude can be written as
\cite{MPS11higgs}
\begin{equation}
V_{\lambda_q\lambda_{\bar{q}}}^{c_1c_2}(q_1,q_2,k_1,k_2)\equiv
n^+_{\mu}n^-_{\nu}V_{\lambda_q\lambda_{\bar{q}}}^{c_1c_2,\,\mu\nu}
(q_1,q_2,k_1,k_2),
\nonumber
\end{equation}
The tensorial part of the amplitude is obtained from Feynman diagrams:
\begin{equation}
\begin{array}{lll}
V_{\lambda_q\lambda_{\bar{q}}}^{\mu\nu}(q_1, q_2, k_1, k_2)
= g_s^2 \,\bar{u}_{\lambda_q}(k_1)
\biggl(\gamma^{\nu}\frac{\hat{q}_{1}-\hat{k}_{1}-m}
{(q_1-k_1)^2-m^2}\gamma^{\mu}-\gamma^{\mu}\frac{\hat{q}_{1}
-\hat{k}_{2}+m}{(q_1-k_2)^2-m^2}\gamma^{\nu}\biggr)v_{\lambda_{\bar{q}}}(k_2).
\end{array}
\end{equation}
The coupling constants $g_s^2 \to g_s(\mu_{r,1}^2)
g_s(\mu_{r,2}^2)$. In the present calculation we take the
renormalization scale to be $\mu_{r,1}^2=\mu_{r,2}^2=M_{q \bar q}^2$. 
The exact matrix element is calculated numerically. Analytical formulae 
are shown explicitly in \cite{MPS11higgs}. 

The off-diagonal parton distributions (i=1,2) are calculated as
\begin{equation}
\begin{array}{lll}
f_i^{\mathrm{KMR}}(x_i,Q_{i,t}^2,\mu_i^2,t_i)  =  R_g
\frac{d[g(x_i,k_t^2)S_{1/2}(k_{t}^2,\mu_i^2)]}{d \log k_t^2} |_{k_t^2
= Q_{it}^2} \;
F(t_i) \; ,
\end{array}
\label{KMR-off-diagonal-UGDFs}
\end{equation}
where $S_{1/2}(q_t^2, \mu^2)$ is a Sudakov-like form factor relevant
for the case under consideration. 
It is reasonable to take the factorization scale as: $\mu_1^2 =
\mu_2^2 = M_{q \bar q}^2$.

The factor $R_g$ here cannot be calculated from first principles
in the most general case of off-diagonal UGDFs.
It can be estimated in the case of off-diagonal collinear PDFs
when $x' \ll x$ and $x g = x^{-\lambda}(1-x)^n$.
Typically $R_g \sim$ 1.3 -- 1.4 at the Tevatron energy. 
The off-diagonal form factors are parametrized here as 
$F(t) = \exp \left( B_{\mathrm{off}} t \right)$.
In practical calculations we take $B_{\mathrm{off}}$ = 2 GeV$^{-2}$.
In evaluating $f_1$ and $f_2$ needed for calculating the amplitude
(\ref{amplitude}) we use different collinear distributions.
%
%

%
%

\section{Results}

In our published papers \cite{MPS10higgs,MPS11higgs} we have calculated 
differential cross sections
not only for exclusive Higgs boson production but also for $b \bar b$ 
and digluon $g g$ production. In all our calculations we take into account off-shellness
of the gluons initiating a relevant hard subprocess. The details about 
the off-shell matrix element for Higgs boson production can be found in
Ref.~\cite{PTS_ggH_vertex}. In contrast to the exclusive
production of $\chi_c$ mesons \cite{PST_chic}, here due to a large 
factorization scale $\sim M_H$, the off-shell effects for 
$g^*g^*\to H$ give only a few percent change of the cross section
compared to the calculation with on-shell matrix elements used in 
the literature.
We use the same unintegrated gluon distributions for Higgs, continuum 
$b \bar b$ and digluon production.

The Higgs boson differential cross sections are calculated assuming
a three-body process $p p \to p H p$.
Assuming full coverage for outgoing protons
we construct two-dimensional distributions 
$d \sigma / dy d^2 p_t$ in Higgs rapidity and transverse momentum. 
The distribution is used then in a simple Monte Carlo code 
which includes the Higgs boson decay into the $b {\bar b}$ channel. 
It is checked subsequently whether $b$ and $\bar b$
enter into the region spanned by the central detector. 

\begin{figure}
\includegraphics[width=6.0cm]{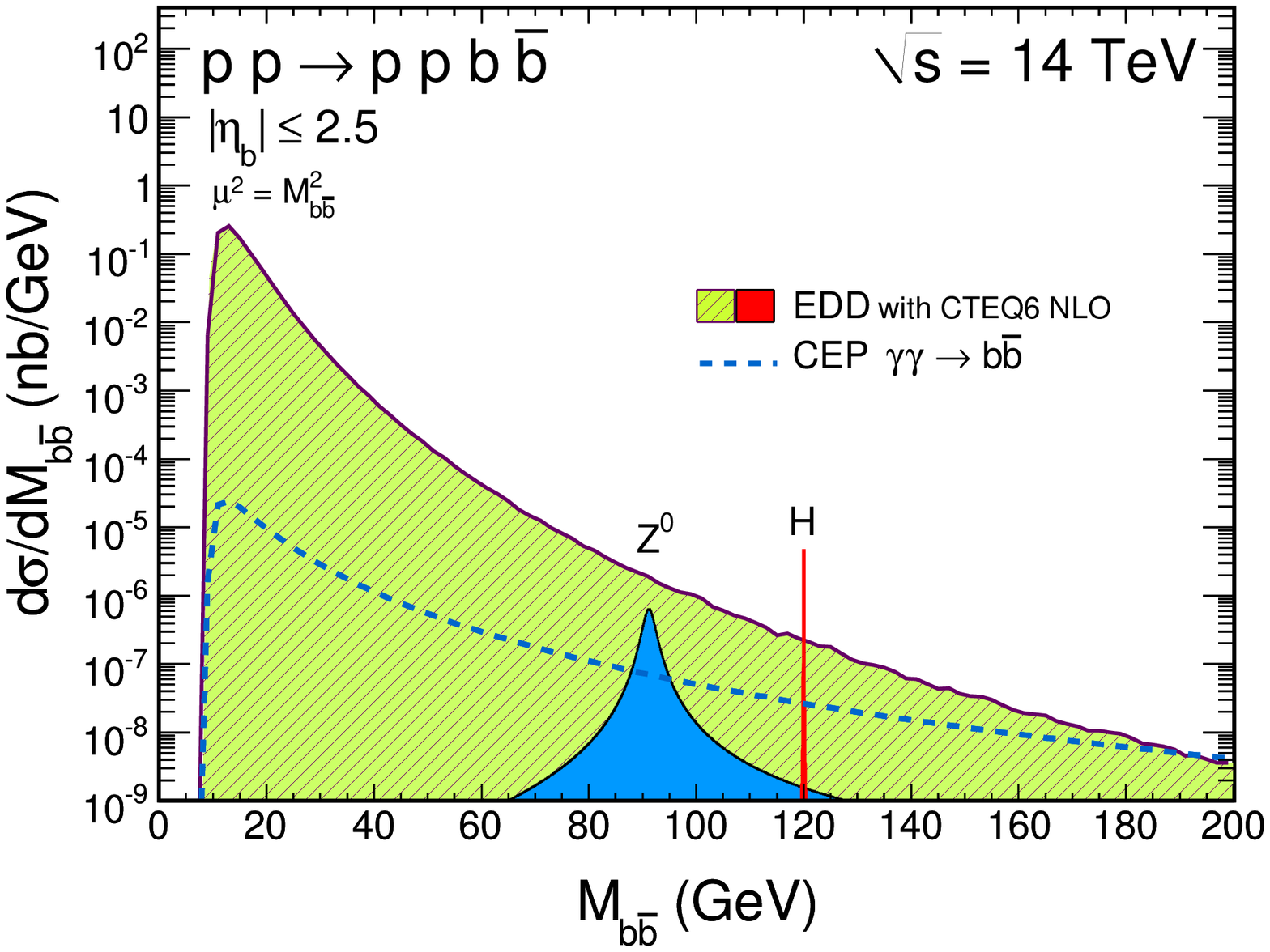}
\includegraphics[width=6.0cm]{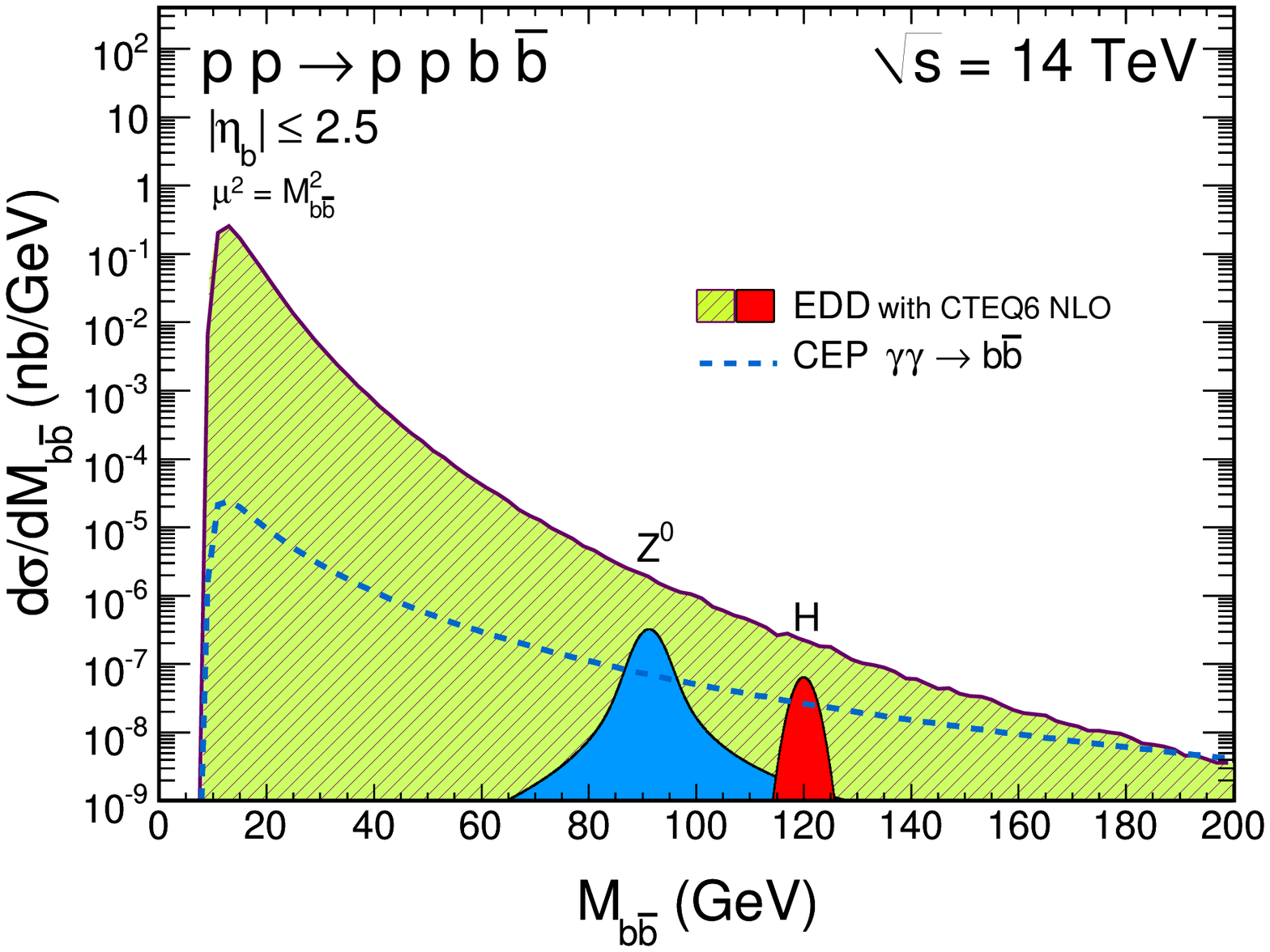}
\caption{The $b \bar b$ invariant mass distribution for $\sqrt{s}$ =
14 TeV and for $b$ and $\bar b$ jets for $-2.5
< \eta < 2.5$ corresponding to the ATLAS and CMS detectors. 
The left panel shows purely theoretical predictions, while the right
panel includes experimental effects due to experimental uncertainty
in missing mass measurement.} 
\label{fig:dsigma_dMbb_fully}
\end{figure}

In the left panel of Fig.\ref{fig:dsigma_dMbb_fully} we show the 
double diffractive contribution for the CTEQ6 \cite{CTEQ} 
collinear gluon distribution and the contribution from 
the decay of the Higgs boson 
including decay width calculated as in 
Ref.~\cite{Passarino_decay_width}, see
the sharp peak at $M_{b \bar b}$ = 120 GeV.
The phase space integrated cross section for the Higgs production,
including absorption effects with gap survival probability $S_G = 0.03$ 
is less than 1 fb.
The result shown in Fig.\ref{fig:dsigma_dMbb_fully} includes 
branching fraction for BR($H \to b \bar b) \approx$ 0.8 
and the rapidity restrictions. The much broader 
Breit-Wigner type peak to the left of the Higgs signal corresponds to 
the exclusive production of the $Z^0$ boson with the cross section 
calculated as in Ref.~\cite{CSS09}.
The branching fraction BR($Z^0 \to b \bar b) \approx$ 0.15 
has been included in addition. In contrast to the
Higgs case the absorption effects for the $Z^0$ production are much
smaller \cite{CSS09}. The sharp peak
corresponding to the Higgs boson clearly sticks above the
background. 

\begin{figure}
\includegraphics[width=6.0cm]{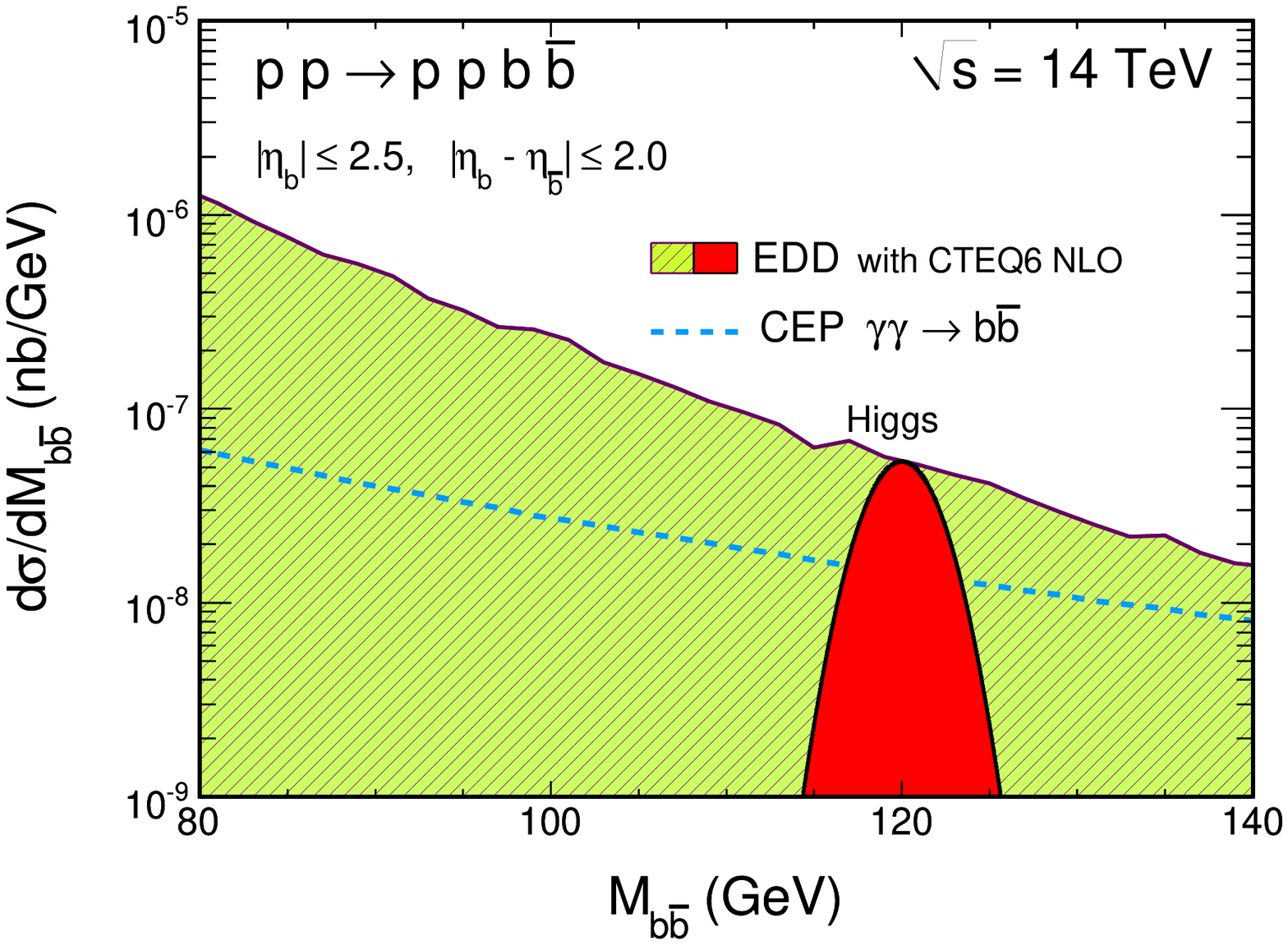}
\includegraphics[width=6.0cm]{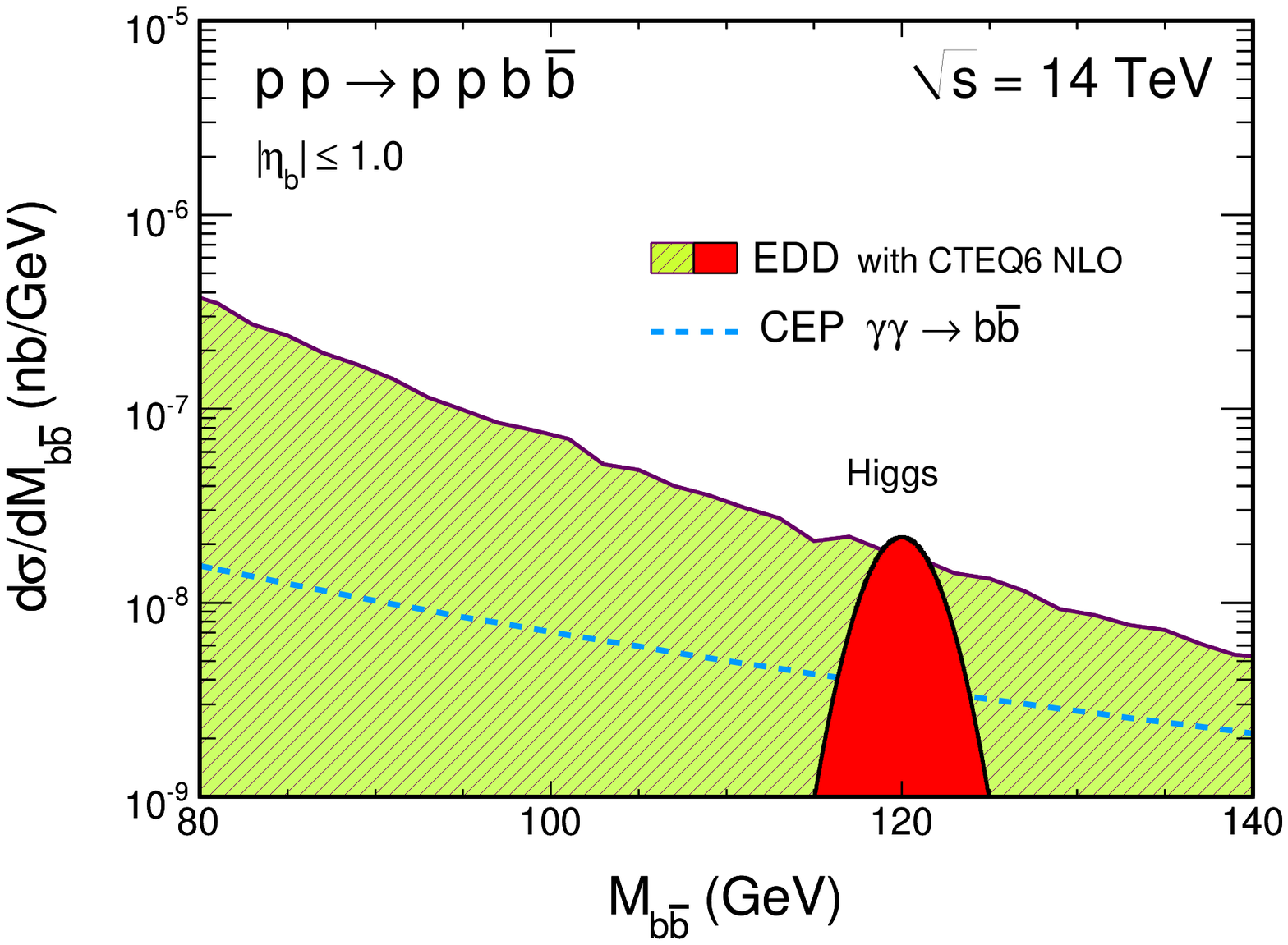}
\caption{The $b \bar b$ invariant mass distribution for $\sqrt{s}$ =
14 TeV. In the left panel the rapidity
difference is limited to $(-1,1)$ and in the right panel both pseudorapidities
are restricted to -1$ < \eta < $1.}
\label{fig:dsigma_dMbb_cuts}
\end{figure}

In reality the situation is much worse as both protons 
and  $b$ and $\bar b$ jets are measured with a certain
precision which automatically leads to a smearing in $M_{b \bar b}$ .
Experimentally instead of $M_{b \bar b}$ one will measure rather
two-proton missing mass. 
In our calculations the experimental effects are included in the
simplest way by a convolution  of the theoretical distributions with 
the Gaussian smearing function with $\sigma$ = 2 GeV
which is due to the precision of measuring forward protons.
In the right panel we show the two-proton missing mass distribution 
when the experimental smearing is included. 
Now the bump corresponding to the Higgs boson is below the $b \bar b$ 
background. 
The situation for some scenarios beyond the Standard Model may be better.

In Refs.\cite{MPS10higgs,MPS11higgs} we have discussed in great detail
how to improve the difficult situation. Examples are shown in 
Fig.\ref{fig:dsigma_dMbb_cuts}. In the left panel we show the situation
when a cut on difference of pseudorapidities is limited to the interval (-1,1)
and in the right panel when cuts on pseudorapidity of $b$ and $\bar b$ are 
imposed.
In both cases the situation seems much better than in the previous
case. We have checked, however, that this is an optimal situation and 
further imporovement of the signal-to-background ratio is in practice 
impossible.

\begin{figure}[!h]
\begin{center}
 \includegraphics[width=8cm]{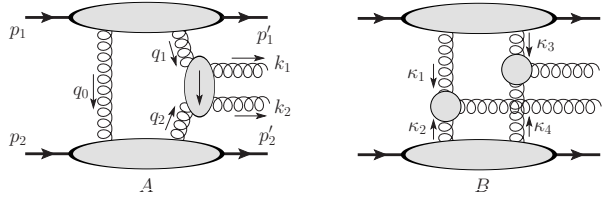}
\end{center}
   \caption{
\small The mechanisms of the digluon production.
 }
 \label{fig:digluon}
\end{figure}

\begin{figure}[!h]
\begin{center}
 \includegraphics[width=6cm]{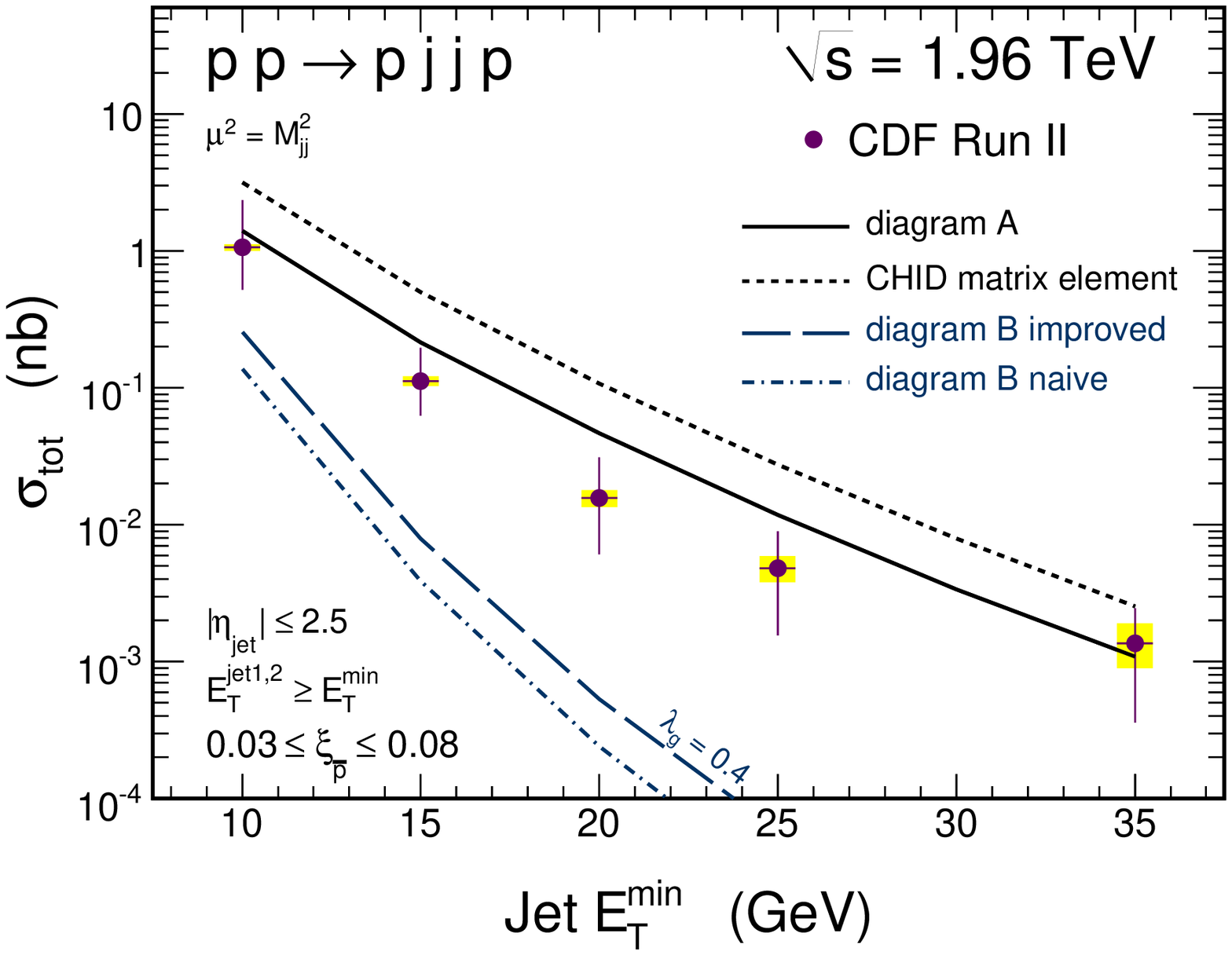}
 \includegraphics[width=6cm]{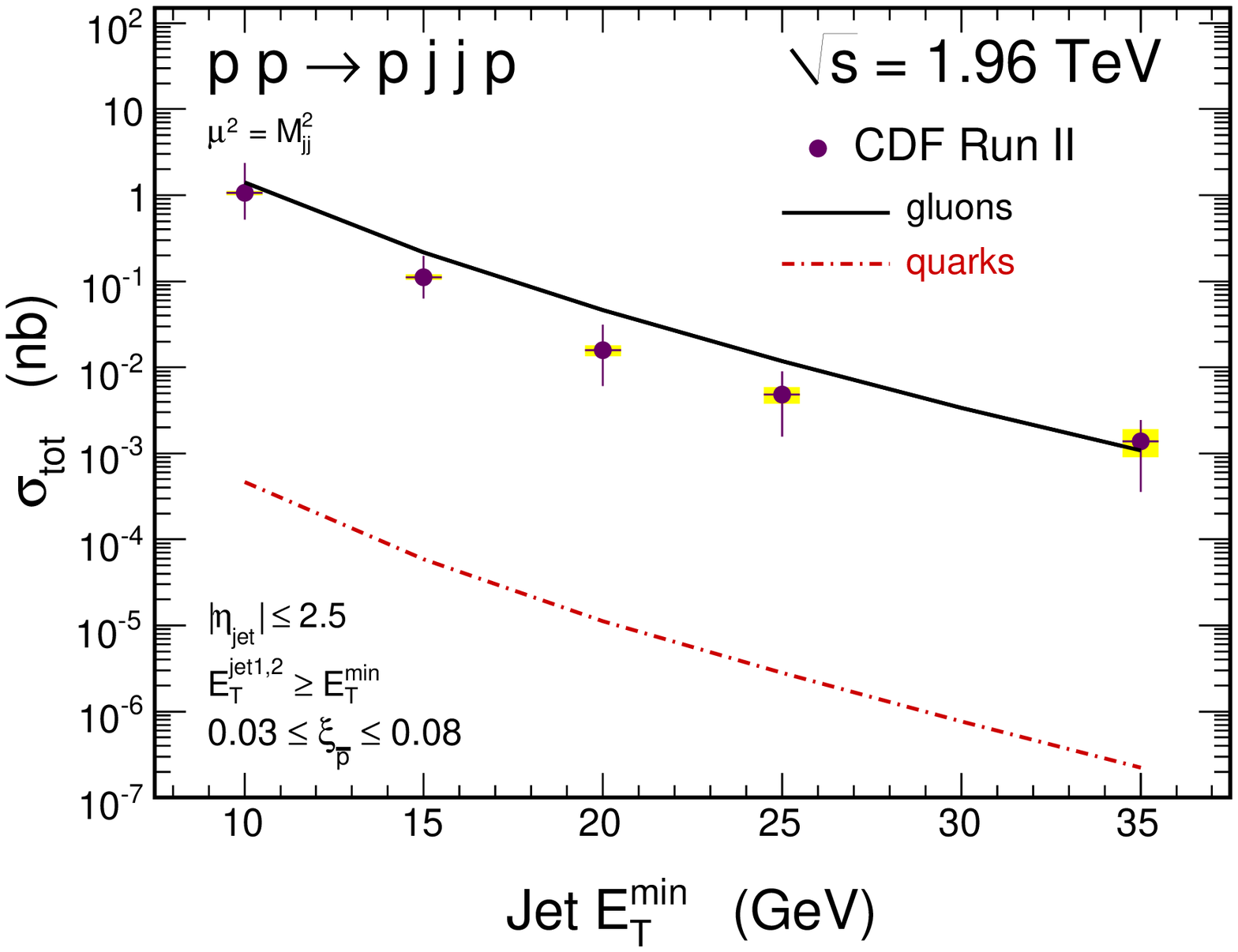}
\end{center}
   \caption{
\small The total cross section for exclusive dijet production as a 
function of $E_{t,min}$. The experimental data points are taken from 
Ref.~\cite{CDF-dijets}. Left panel: digluon contribution for the standard 
mechanism with our matrix element
(solid line) and CHID matrix element (short-dashed line), for
diagram B (long-dashed line). Right panel: quark-antiquark
(dash-dotted line) contribution. }
 \label{fig:cdf_etmin}
\end{figure}

Now we come to the distributions for exclusive dijet production. 
The diagrams of the possible mechanisms are shown in 
Fig.\ref{fig:digluon}. The first diagram is the dominant mechanism
of dijet production, whereas the second mechanism was discussed
in more detail in our recent paper \cite{MPS2011_gluons}.
The details of the corresponding calculations can be found 
in \cite{MPS2011_gluons}. In Fig.\ref{fig:cdf_etmin}
we show the total cross section as a function of minimal $E_T$.
Already the digluon contribution (thick solid line) is slightly above the data.
In the case of quark-antiquark dijets we present the contribution of $u
\bar u, d \bar d, s \bar s, c \bar c$ and $b \bar b$. In the first
three cases, we put the quark masses to zero, and in the last two
cases we take explicit masses known from the phenomenology (1.5 GeV
and 4.75 GeV, respectively). The sum of all quark-antiquark
contributions is shown in the right panel by the dash-dotted curve.
We conclude that the quark-antiquark jet contribution is smaller by
more than two orders of magnitude than the digluon one. However, 
the $b \bar b$ contribution can be essential e.g. as a background for 
Higgs searches in exclusive $pp$ scattering.

The exceptional dominance of digluon jets over quark-antiquark jets
found here offers a extraordinary conditions for increased glueball production
in gluon fragmentation \cite{MPS2011_gluons}. In order to investigate it more 
one needs to study a contamination of central diffractive components where
the proportions of digluonic to quark-antiquark are less favourable.

\begin{figure}[h!]
\begin{center}
\includegraphics[width=7.0cm]{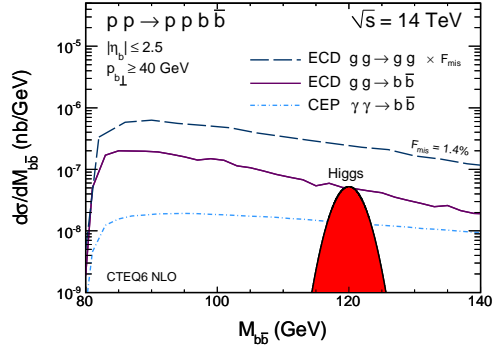}
\end{center}
 \caption{
Invariant mass distribution of the $b \bar b$ system. Shown are
contributions from diffractive Higgs boson (shaded area), $b \bar b$ 
continuum (solid line), $\gamma \gamma$ continuum (dash-dotted line) and
diffractive digluon contribution (dashed line) multiplied by the ATLAS
misidentification factor squared.
}
\label{fig:dsig_dMjj_Higgs}
\end{figure}

The gluonic jets can be misidentified as $b$-quark jets.
For example the ATLAS misidentification factor is 1.4\%.
If both gluonic jets are misidentified then such a misidentified event 
can contribute to a background to exclusive Higgs boson production.
In Fig.\ref{fig:dsig_dMjj_Higgs} we illustrate the situation.
We show both the Higgs signal (hatched area) including experimental
resolution  as well as diffractive 
$b \bar b$ continuum, QED $b \bar b$ continuum as well as formally
reducible digluon contribution. In the calculation we have assumed
that jet misidentification probability is 1.4\%, 
i.e. we have multiplied the dijet cross section by a small number 
0.014$^2$.
The obtained digluon contribution is even larger than the $b \bar b$ one and
overlays the Standard Model Higgs signal. In the case of 
Minimal Supersymmetric Model the situation can be more favourable.
\cite{Cox:2007sw,HKRSTW08}.

\section{Conclusions}

We have shown and discussed differential distributions for the continuum
$b \bar b$ production. Inclusion of the experimental resolution is 
necessary when comparing the Higgs signal and the $b \bar b$ background.
Our analysis shows that a special cuts could be useful to see a
signal above the continuum background. We have also shown a reducible 
background due to a misidenification of gluonic jets as $b$ or $\bar b$
jets.

Rough agreement of the theoretical dijet cross section with the Tevatron
data gives more confidence to the predictions for exclusive Higgs
boson production.

Our analysis indicates that a real experiment for the exclusive Higgs
boson production can be rather difficult.
The situation could be better for some scenarios beyond the
Standard Model.

\vspace{-0.3cm}
 
\section*{Acknowledgements} 
I am indebted to Rafa{\l} Maciu{\l}a and Roman Pasechnik
for collaboration on the issues presented here.


\end{document}